\documentclass[preprint,12pt,authoryear, review]{elsarticle}
\usepackage{graphicx}
\usepackage{subfig}
\usepackage[utf8]{inputenc}
\usepackage{bm}
\usepackage[ruled,vlined]{algorithm2e}
\usepackage{amsfonts}
\usepackage{enumerate}
\usepackage{amssymb} %% The amssymb package provides various useful mathematical symbols
\usepackage{amsmath}
\usepackage{amsfonts}
\usepackage{bm}

\newpageafter{abstract} %%Start first section on new page

\begin{document}

\begin{frontmatter}

\title{Stochastic uncertainty analysis of gravity gradient tensor components and their combinations}
\renewcommand{\thefootnote}{\fnsymbol{footnote}}

\author{Pejman Shamsipour\footnotemark[1]\footnotemark[2]\footnotemark[5], Amin Aghaee\footnotemark[1]\footnotemark[4]\footnotemark[5], Tedd Kourkounakis\footnotemark[1]\footnotemark[3]\footnotemark[5],  and Shawn Hood\footnotemark[1]\footnotemark[4]\footnotemark[5]}

\address{
{pejman, amin, tedd.kourkounakis, shawn} @goldspot.ca\footnotemark[1];\\
GoldSpot Discoveries Corp., Montreal, Quebec 980 Rue Cherrier, Suite 201 H2L 1H7 \footnotemark[2];   \\
GoldSpot Discoveries Corp., Toronto, Ontario 69 Yonge Street, Suite 1010 M5E 1K3 \footnotemark[3];\\
GoldSpot Discoveries Corp., 303- 700 West Pender, Vancouver, BC, V6A 1V7 \footnotemark[4];

All authors share equal contribution towards this work.\footnotemark[5]

}

\begin{abstract}
Full tensor gravity (FTG) devices provide up to five independent components of the gravity gradient tensor. However, we do not yet have a quantitative understanding of which tensor components or combinations of components are more important to recover a subsurface density model by gravity inversion. This is mainly because different components may be more appropriate in different scenarios or purposes. Knowledge of these components in different environments can aid with selection of optimal selection of component combinations. 
In this work, we propose to apply stochastic inversion to assess the uncertainty of gravity gradient tensor components and their combinations. The method is therefore a quantitative approach. The applied method here is based on the geostatistical inversion (Gaussian process regression) concept using cokriging. The cokriging variances (variance function of the GP) are found to be a useful indicator for distinguishing the gravity gradient tensor components. This approach is applied to the New Found dataset to demonstrate its effectiveness in real-world applications.
\end{abstract}

\begin{keyword}
geostatistics \sep Gaussian process \sep uncertainty analysis \sep gravity gradient tensor
\end{keyword}

\end{frontmatter}

\section{Introduction} 
Gravity gradiometry permits single components and combinations of components to be applied in the interpretation and inversion of data. The effectiveness of choosing either the single components or the combinations highly depends on the information content of the components and their combinations. 
a structure's trends, noise, orientation, among others, can significantly effect which components are to be selected. As there are a wide range of combinations and more are likely to be suggested in the future, a quantitative ranking of these values is vital.

The majority of quantitative analyses of tensor components were limited to the comparison of gravity measurements and gradiometry gradient tensor \citep{murphy2007target, jordan1978, vasco1989}.

In other approaches, authors attempted to compare the 3D inversion results with known geology qualitatively (\cite{martinez2013d, zhdanov2004, li20013}).
Previous works \citep{pilkington2012analysis, pilkington2013evaluating, mikhailov2007tensor, beiki2010eigenvector} compared tensor components and their combinations in terms of their information content and error levels when inverting for simple models. In these cases the models used were either very simple or did not include the effects of full covariance errors. Here, we use cokriging to study different tensor components using more complex and realistic models.

Commonly, cokriging is used as a mathematical interpolation and extrapolation tool that uses the spatial correlation between primary and secondary variables to improve the estimation of the primary variable. However, in this study cokriging can be seen as an inversion method \citep{Shamsipour2010, tchikaya20163d}. The benefit of performing inversion using cokriging is that we can use the cokriging error as an indicator to determine the uncertainty of estimated densities obtained from different gravity gradients or any combination of them. Even though the cokriging error does not directly depend on the data values used for estimation, it can be used as an  indicator of uncertainty.

The cokriging variance is available for each prism, so we can determine which component gives a better density estimation at each individual location. After testing, we realized that the cokriging variance for different components or their combinations is distinguishable. In order to investigate the effectiveness of different components, we first assume a geometry with fixed covariance parameters. The only difference for the different components is the kernel, or Green's function. However it is possible to have different geometries by changing the parameters of the covariance (lengthscales/ranges and rotation angles) and investigate which component gives the better estimate for each geometry. We can also consider the effect of the noise in each case by adding a nugget effect to the covariance model. To validate our experiments, we conduct stochastic inversion on gravity gradiometer data provided in the New Found dataset. 

%%---------------------------------------------Methodology-------------------------------------------
\section*{Methodology}
%\subsection{Forward modeling}
\subsection{Stochastic inversion and error of estimated model}
\cite{Shamsipour2010} presents the application of cokriging to invert potential field data. The method can be applied to any linear geophysical problem. Consider that there are $n$ data observations $d$ and $m$ parameters (physical properties) on rectangular prisms ($p$), their relationship can be written in matrix form:
\begin{equation}
\label{eq:Gro}
d_{n\times 1} =A_{n\times m}p_{m\times 1}
\end{equation}
where $A$ is the matrix of the geometric terms (kernel).  Since $d$ and $p$ are linearly related, then, their covariance matrices are also linearly related:
\begin{equation}
C_{dd}=AC_{pp}A^T+C_0
\label{eq:th}
\end{equation}
where $C_{dd}$ and $C_{pp}$ are respectively the $n \times n$ and $m \times m$ covariance matrices for observations and parameters, and $C_0$ is the $n \times n$ data error covariance matrix which is usually modeled as a nugget effect (white noise). We also have:
\begin{equation}
C_{dp}=AC_{pp}
\label{eq:cross}
\end{equation}
where $C_{dp}$ is the $n \times m$ cross-covariance matrix between the observation data and parameters. Finally, the estimates of parameters (primary variable) are obtained from the observation data (secondary variable):
\begin{equation} 
p^*=\Lambda^{T}d
\label{eq:xstar}
\end{equation}
where $\Lambda$ is the optimal matrix of weighting coefficients obtained by minimization of the estimation variance. The estimation variances are found on the diagonal of the following matrix \citep{myers1982}:
\begin{equation}
E((p-p^{*})(p-p^{*})^T)=C_{pp}-2\Lambda^{T}C_{dp}+\Lambda^{T}C_{dd}\Lambda 
\end{equation}
where $\Lambda$ is the matrix of weighting coefficients. Here $p$ and $d$ are multidimensional random variables. Minimization of the above estimation variance yields the simple cokriging solution
\begin{equation}
C_{dd}\Lambda=C_{dp}.
\end{equation}
Finally, the vector of cokriging variances is obtained from:

\begin{equation} 
\sigma_{ck}^2=diag(C_{pp}-\Lambda^TC_{dp})
\label{eq:ckv}
\end{equation}
where diag(.) finds the diagonal elements of a matrix and put them in a vector. Note that the maximum variance equals the overall variance ($C_{pp}$) and when the weights increase, this variance drops.

\subsection{Gaussian processes for inversion and error estimation}
Gaussian processes are non-parametric priors and a common machine learning framework that have been applied to both regression and classification tasks (\cite{rasmussen2003gaussian}). Through this analysis we will observe the equivalence between cokriging inversion within the geostatistics domain and Gaussian process inversion.

Bayesian inversion corresponds to computing the posterior distribution of the parameters of interest $\phi$ given our observations $y$.  \cite{reid2013bayesian} provide an overview of the application of inversion within Gaussian processes. Given an $n \times m$ sensitivity matrix, $A$, and $K_{\phi\phi}$ as the $m \times m$ covariance matrix, we may represent the covariance structures of our observations $y$ and parameters $\phi$  as:

\begin{equation}
K_{yy}=AK_{\phi\phi}A^{T} + \sigma^{2}I
\label{eq:gp_th}
\end{equation}
\begin{equation}
K_{y\phi}^{T}=AK_{\phi\phi}
\label{eq:gp_cross}
\end{equation}

where $\sigma^{2}$ is the noise variance. These representations are identical to those of equations \ref{eq:th} and \ref{eq:cross}, respectively. From here, a predictive distribution can represented by a Gaussian  prior as

\begin{equation}
\phi|A,y \backsim \mathcal{N} (\phi|\mu_{\phi|y}, \Sigma_{\phi|y})
\end{equation}
with
\begin{equation}
\mu_{\phi|y} = K_{y\phi}^{T}K_{yy}^{-1}y
\label{eq:gp_xstar}
\end{equation}
\begin{equation}
\Sigma_{\phi|y} = K_{\phi\phi} - K_{y\phi}^{T}K_{yy}^{-1}K_{y\phi}
\label{eq:gp_ckv}
\end{equation}

Note that the matrix of weighting coefficients $\Lambda$ introduced in equation \ref{eq:xstar} is similarly represented by $K_{y\phi}^{T}K_{yy}^{-1}$. The cokriging variance vector $\sigma_{ck}^{2}$ of equation \ref{eq:ckv} could be calculated as $diag(\Sigma_{\phi|y})$. Through these comparisons, we can clearly visualize the similarities between Gaussian process inversion and the cokriging inversion used in this work.

\subsection{Forward modeling}
The vertical component of the gravitational effect of a right rectangular prism with density $\rho$; $x$ limits $a_1$, $a_2$; $y$ limits $b_1$, $b_2$; and $z$ limits $c_1$ and $c_2$ is
\begin{equation}
g^p=-\gamma\rho \sum_{i=1}^{2}\sum_{j=1}^{2}\sum_{k=1}^{2}\mu_{ijk}\left[ x_i ln(y_i+r_{ijk})+ln y_j(x_i+r_{ijk})-z_k arctan \frac{x_i y_j}{z_k r_{ijk}}\right] 
\label{eq:gref}
\end{equation}
where $\gamma$ is the Newton’s gravitational constant, and
\begin{equation}
x_i=x-a_i, y_j=y-b_j, z_k=z-c_k
\end{equation}
\begin{equation}
r_{ijk}=\sqrt{x_i^2+y_j^2+z_k^2}
\end{equation}
\begin{equation}
\mu_{ijk}=(-1)^{i+j+k}.
\end{equation}

The tensor components for $g$ in equation (\ref{eq:gref}) are given by (Forsberg, 1984; Li and Chouteau, 1998)
\begin{equation}
T_{xx}^p=\gamma\rho \sum_{i=1}^{2}\sum_{j=1}^{2}\sum_{k=1}^{2}\mu_{ijk} arctan \frac{y_j z_k}{x_i r_{ijk}}, 
\label{eq:Txx}
\end{equation}
\begin{equation}
T_{xy}^p=-\gamma\rho \sum_{i=1}^{2}\sum_{j=1}^{2}\sum_{k=1}^{2}\mu_{ijk} ln(z_k+r_{ijk}),
\label{eq:Txy}
\end{equation}
\begin{equation}
T_{xz}^p=-\gamma\rho \sum_{i=1}^{2}\sum_{j=1}^{2}\sum_{k=1}^{2}\mu_{ijk} ln(y_i+r_{ijk}),
\label{eq:Txz}
\end{equation}
\begin{equation}
T_{yy}^p=\gamma\rho \sum_{i=1}^{2}\sum_{j=1}^{2}\sum_{k=1}^{2}\mu_{ijk} arctan \frac{x_i z_k}{y_j r_{ijk}}, 
\label{eq:Tyy}
\end{equation}
\begin{equation}
T_{yz}^p=-\gamma\rho \sum_{i=1}^{2}\sum_{j=1}^{2}\sum_{k=1}^{2}\mu_{ijk} ln(x_i+r_{ijk})
\label{eq:Tyz}
\end{equation}
and
\begin{equation}
T_{zz}^p=\gamma\rho \sum_{i=1}^{2}\sum_{j=1}^{2}\sum_{k=1}^{2}\mu_{ijk} arctan \frac{x_i y_j}{z_k r_{ijk}}, 
\label{eq:Tzz}
\end{equation}
Please note that 
\begin{equation}
T_{xx}^p+T_{yy}^p+T_{zz}^p=0
\label{eq:Txyz}
\end{equation}
which means that out of the six tensor components, only five are independent.

\subsection{Gravity gradients inversion}
Consider that there are n gravity data ($g=[g_1, g_2, ..., g_n]$) and $m$ density values on rectangular prisms ($\rho=[\rho_1, \rho_2, ..., \rho_m]$), their relationship can be written in the matrix form as
\begin{equation}
g=G\rho
\end{equation}
where $G$ is the $n\times m$ kernel and its elements are calculated using the formula in equation (\ref{eq:gref}). For each tensor component of $g$, similar expressions can be written. For example 
\begin{equation}
T_{xx}=G_{xx}\rho
\end{equation}
where the elements of $G_{xx}$ are calculated using the formula in equation (\ref{eq:Txx}). This means that stochastic inversion can be applied on each of the tensor components using the procedure explained in the previous section and error of the estimated model. Similarly, error of the estimated model can be calculated for each tensor component.

Combination of different tensor components can also be used by stochastic inversion. For example for combination of $T_{xx}$ and $T_{xy}$ we have
\begin{equation}
\left[ T_{xx} T_{xy}\right] = 
\begin{bmatrix}
G_{xx}\\
G_{xy}
\end{bmatrix}\rho
\end{equation}
Please note that since based on equation (\ref{eq:Txyz}) out of $T_{xx}$, $T_{yy}$ and $T_{zz}$, only two are independent, and as such only two of them should be used in a combination.

%%--------------------------Results and Discussion--------------------------
\section{Results \& Dibscussion}
\begin{figure}[hbtp]
\centering
\includegraphics[width=1\textwidth]{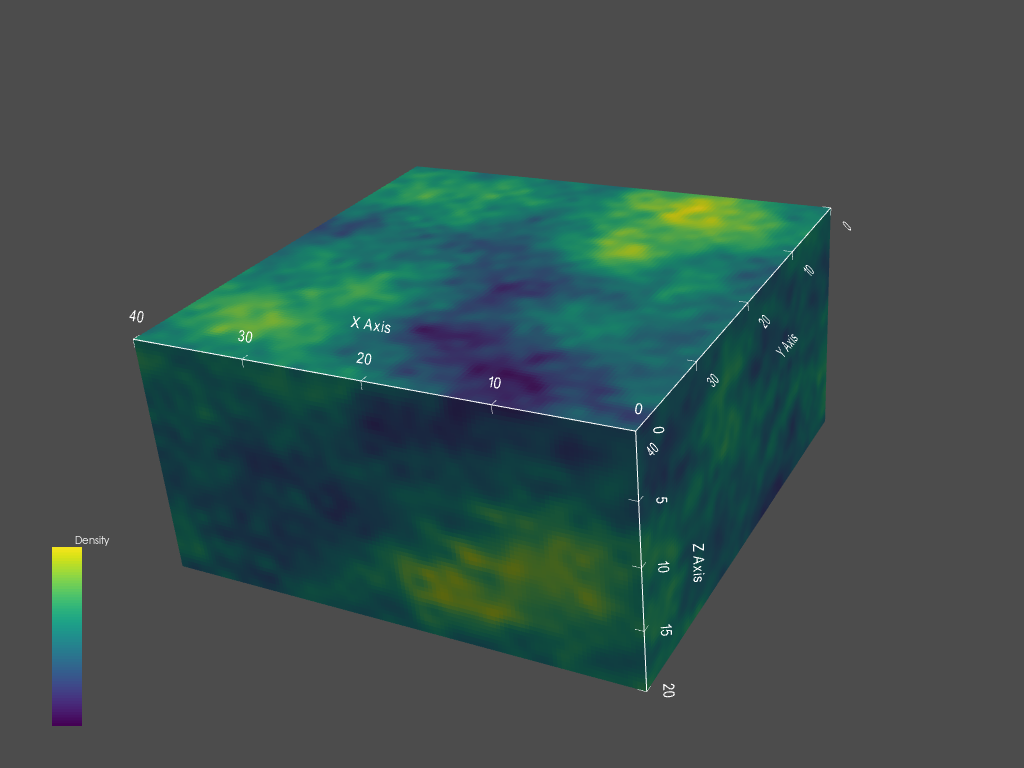}
\caption{Stochastic density model simulated using FFTMA simulation method, with normalized density values. Exponential isotropic variogram with range of 10 m is utilized to generate the model.}
\label{fig:Sto_model}
\end{figure}

\begin{figure*}[]
\centering
\includegraphics[width=1\textwidth]{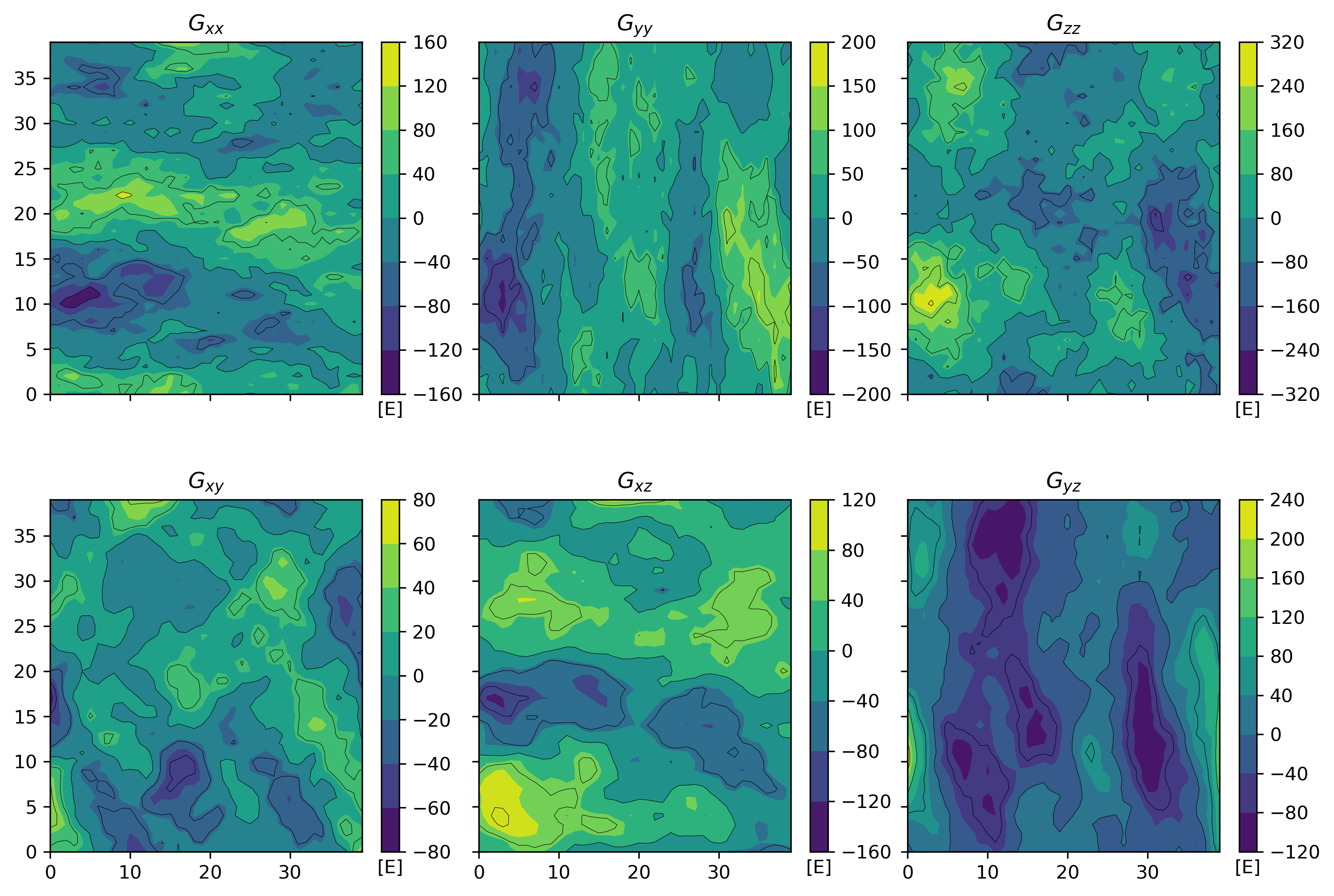}
\caption{Gravity tensor component quantities calculated from the stochastic density model.}
\label{fig:grad_comp}
\end{figure*}

We generate densities for $1\times 1\times 1$ m prisms by FFTMA simulation, which is designed for 3D Gaussian model with variable covariance model properties simulation \citep{le2000fft}. We  applied a spherical variogram model with C$=90000$ kg/m$^3$ and $a_x=a_y=20$ and $a_z=10$ m where C is the variogram sill, and $a_x,  a_y, a_z$ are the variogram ranges in different directions. 
The 3D domain is divided into $40\times 40\times20=32000$ cubic prisms. The stochastic density model is shown in Figure \ref{fig:Sto_model}. Using these density values, we calculate all gravity gradient components (Li and Chouteau (1998)) shown in Figure \ref{fig:grad_comp}. 

\begin{figure*}[]
\includegraphics[ width=.9\textwidth]{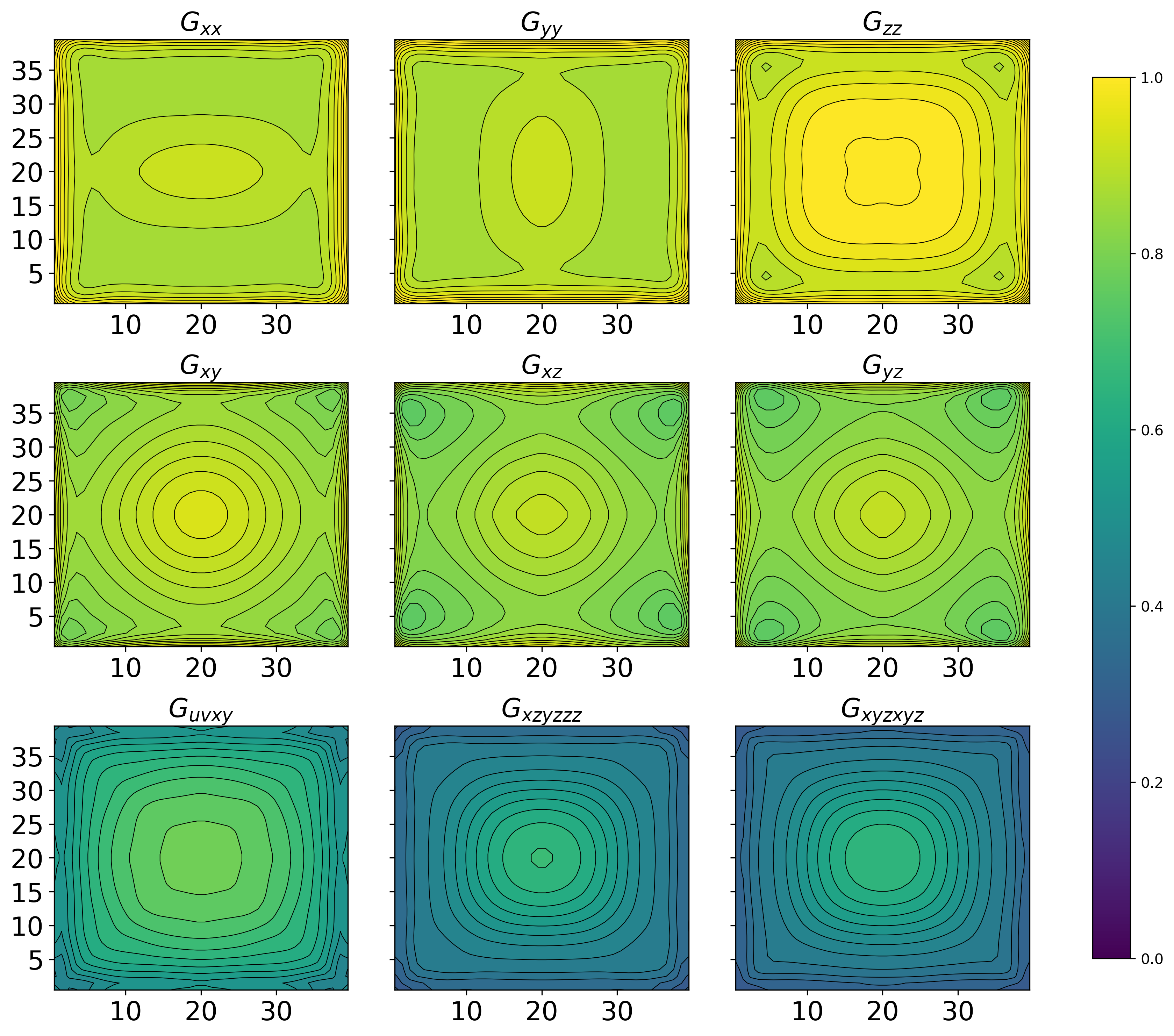}
\caption{Cokriging variances shown in section Z = 5 m depth for six components and three component combinations. The variance values have been normalized.}	
\label{fig:var_iso}
\end{figure*}

\begin{figure*}[]
\includegraphics[ width=.9\textwidth]{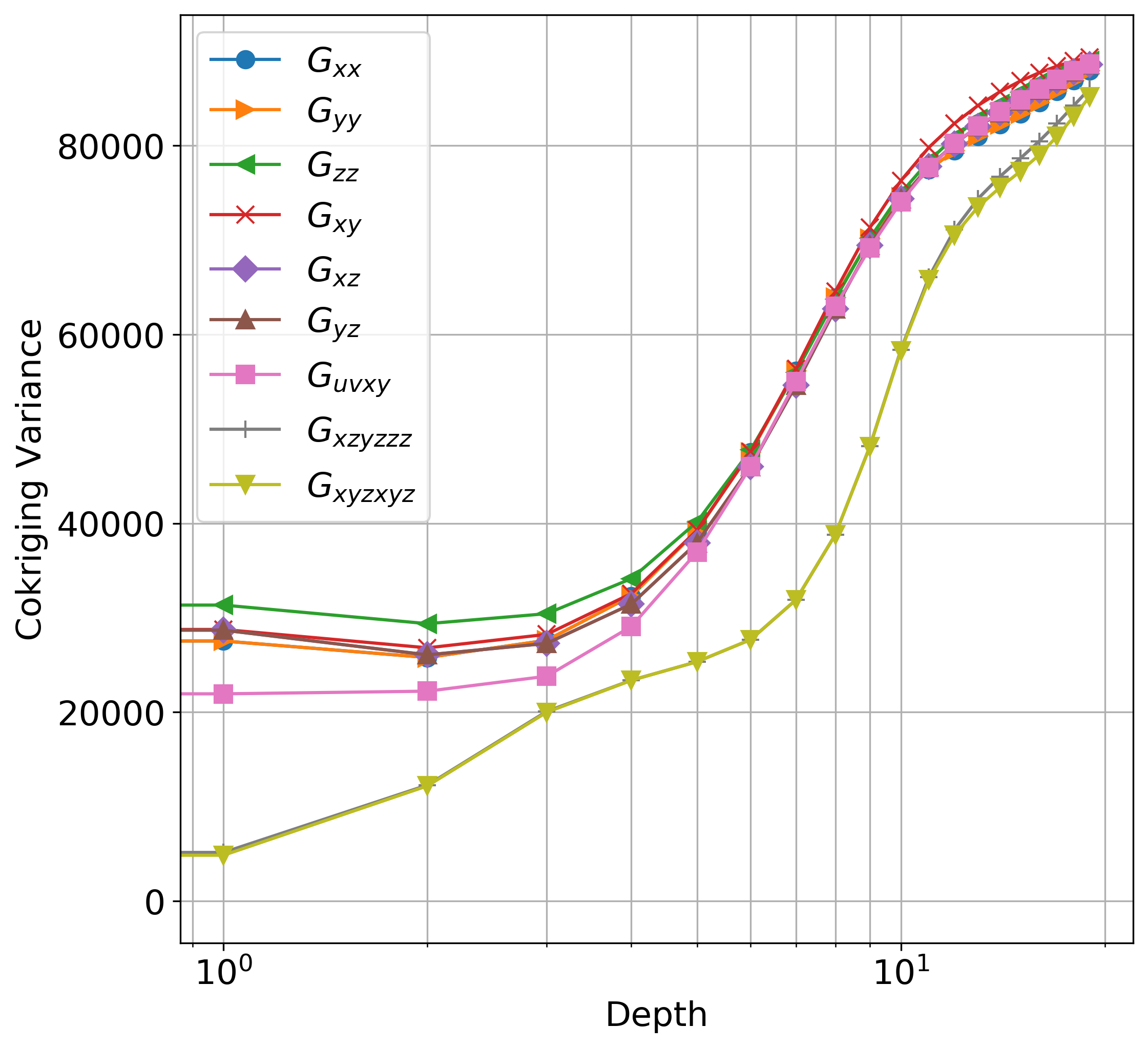}
\caption{Cokriging variances for several components and component combinations, as a function of depth for one cell in centre grid of each layer.}	
\label{fig:cent}
\end{figure*}

Then cokriging variances for each of the components for all prisms are calculated using Eq. \ref{eq:ckv}. These cokriging variances (error estimations) are shown in  Figure \ref{fig:var_iso} for a depth $Z=5$ m. We can easily see the differences between estimation errors for all components. It should be noted the difference of cokriging variances between component layers close to the surface are minimal. However, it is still possible to distinguish between these tensor components. We should also point out that the patterns close to the edges are less liable because of the edge effect.
The cokriging (error) variances for each case have a distinguishable structure. This results suggest that a gradient component, or combination of components, can be chosen lead to a more advantageous result in parameter estimation, depending on the given application. In Figure \ref{fig:cent}, we choose a $3 \times 3$ grid in the center of the first layer below the surface of the horizontal layers in the z-direction. From these, the cokriging variances are drawn for several components and component combinations, as a function of depth for each $3 \times 3$ prism. We can see that each component shares an upward trend in variance as the depth increases, however the $Gxzyzzz$ and $Gxyzxyz$ components show significantly reduced variance as compared to the other component combinations. These findings are even more prominent at shallower depths.

\subsection{Evaluation of Hyperparameters}
We discuss the effects of hyperparameter tuning on the cokriging variances for each kernel component. This includes analysis of the range (lengthscale), as well as rotation angles, $\theta$. Further investigations focus on the effects of manipulating the error covariance, or nugget effect.

\subsubsection{Effects of Range}
\begin{figure*}[]
\centering
\includegraphics[width=.9\textwidth]{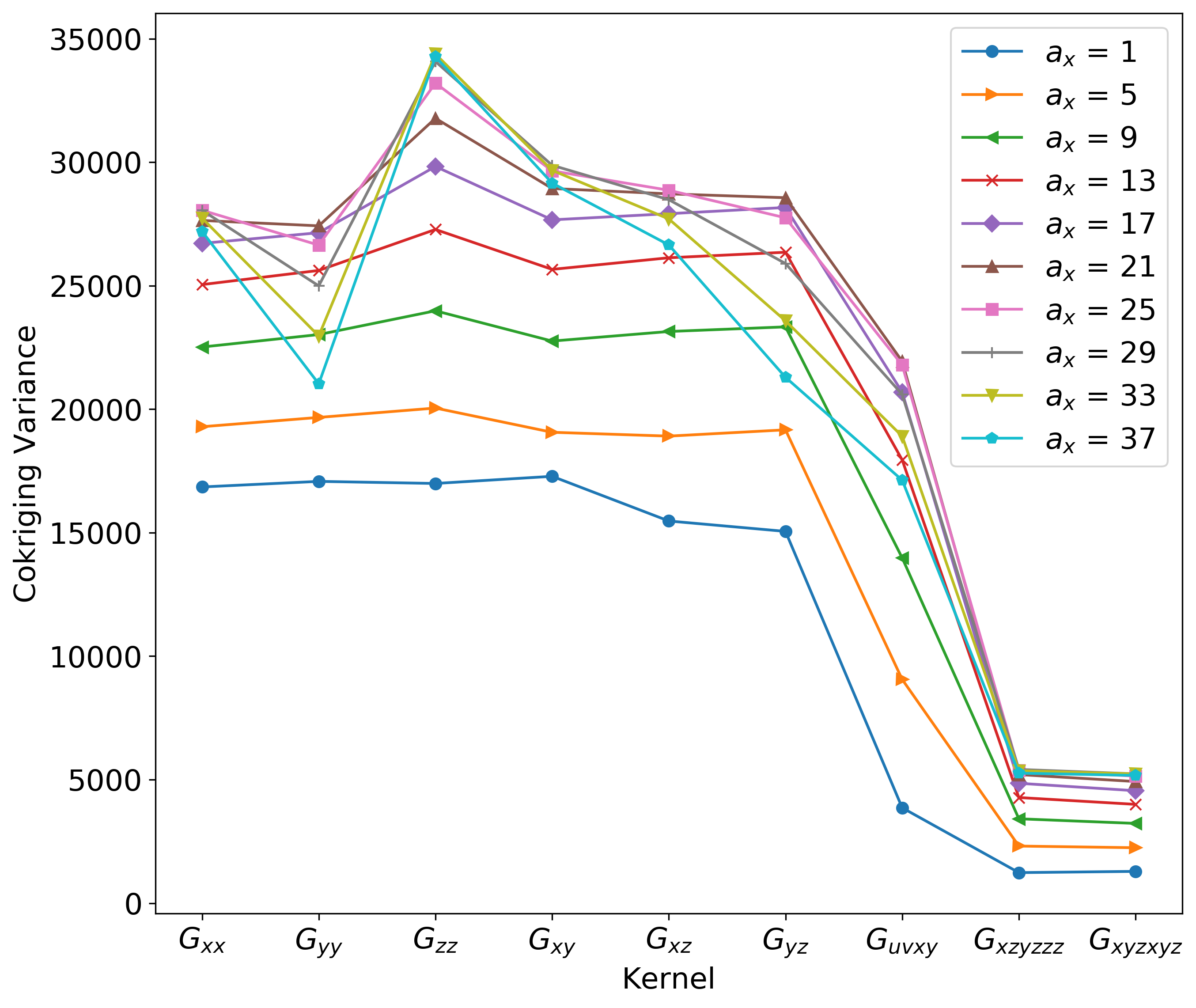}
\caption{Cokriging variances of several components and component combinations as a function of increasing range. In this example, the increasing range is in the x direction.}
\label{fig:geo_ax}
\end{figure*}

In order to investigate more, we select a $3 \times 3$ grid at the center of the first horizontal layer below the surface. We change the variogram range, ax, from 1 meter to 37 meters, and calculate variances for several components and component combinations. The variances for different variogram ranges are shown in Figure \ref{fig:geo_ax}. It is seen that different gradients have different behaviours. For example, if we look at $Gxx$ and $Gyy$ components, at lower ranges they are similar but at the higher ranges $Gyy$ starts to show lower error variance. Note that for the combination of components, the effects of changing of changing range are less significant as compared to single components.

\subsubsection{Geometric Rotation}
\begin{figure*}[]
\centering
\includegraphics[width=.9\textwidth]{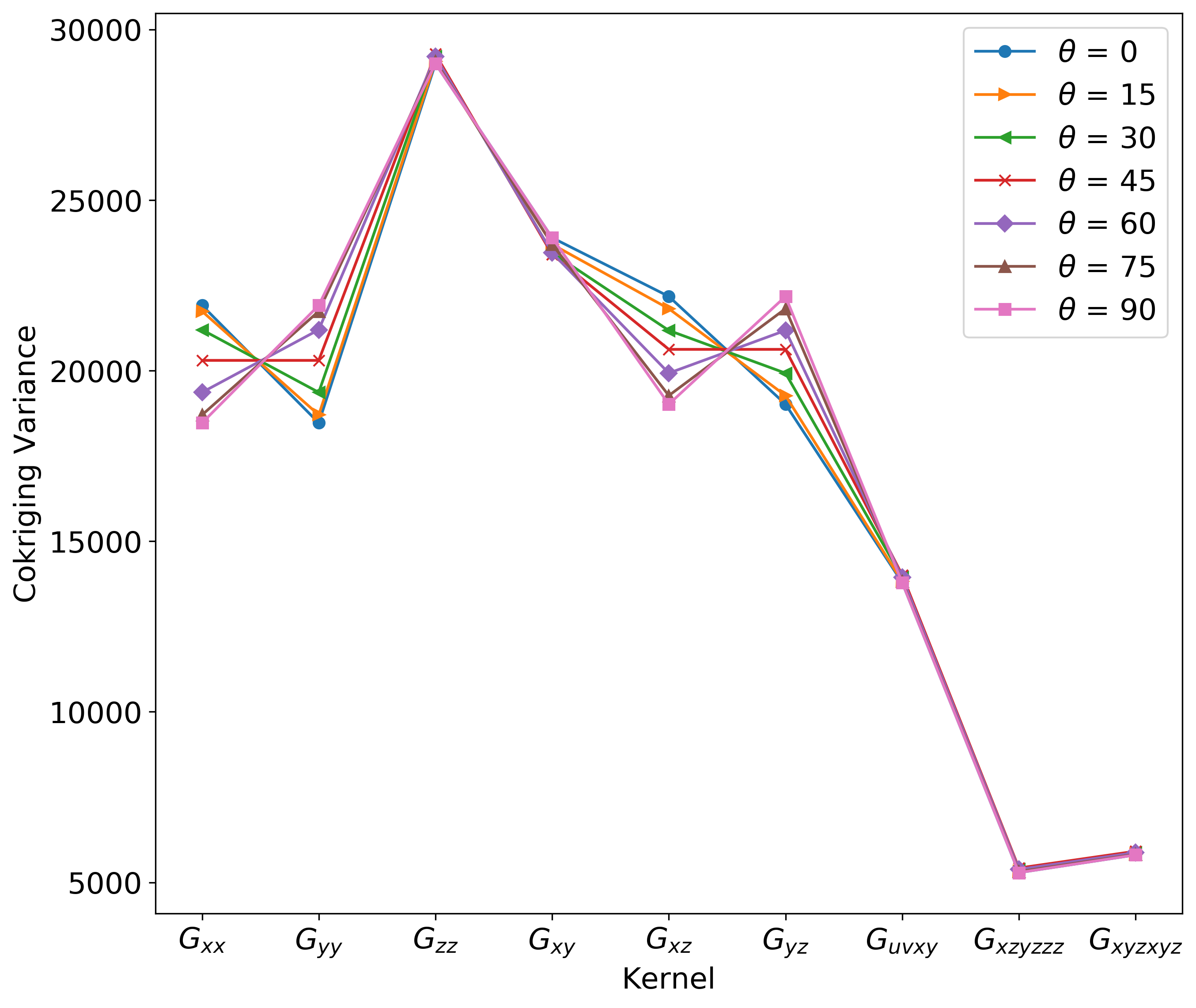}
\caption{Cokriging variances of several components and component combinations as a function of increasing coordinate system rotation angle.}
\label{fig:rotation}
\end{figure*}

To further study the effects of geometry, the outcomes of changing the coordinate system rotation angle, $\theta$, for a given prism have been measured. This experiment has been visualized in Figure \ref{fig:rotation}, where $\theta$ changes about the z-axis at regular intervals between 0 and 90 degrees. The noise of single gravitational tensor component is sensitive to the angle of rotation. As expected, tensor components that do not contain elements of the rotated axis ($Gxx$, $Gyy$, $Gxy$) remain unaffected. This sensitivity is significantly reduced when applied to a greater combination of components. These combinations of tensor components also yield significantly lower variance than any single component.

\subsubsection{Study of Noise}
\cite{switzer1993spatial} discussed how the kriging error reflects two scales of variability:
\begin{itemize}
\item Local modulation: simply indicates the spatial configuration of the data locally. For example, errors are smaller close to data points, clustered data carry less information, and so on.

\item Regional modulation: as we apply a variogram for the whole domain, where data allows,  the variogram parameters are adjusted regionally, and the kriging variance, without being conditional, becomes an useful indicator of uncertainty.
\end{itemize}
In this manuscript we take advantage of both the regional modulation effect and the effect of the different tensor kernels to analyze the errors from gravity gradients.

For the same prism, the effect of adding noise using nugget effect is studied. The nugget effect is a phenomenon present in many regionalized variables and represents short scale randomness or noise in the regionalized variable. It can be seen graphically in the variogram plot as a discontinuity at the origin of the function. 

Before we can continue with these experiments, we must account for any preprocessing that may amplify the noise of any affected  gravitational tensor components. \cite{pilkington2013evaluating} studied the affect of noise and determined that regardless of the preprocessing method and input error, that tensors will maintain a static relative noise as compared to the $Gzz$ component. Using a ratio of 1:0.59:0.37:0.7 for the $C_{0}$ noise levels for the $Gzz$, $Gxx$/$Gyy$, $Gxy$, and $Gxz$/$Gyz$, respectively, the postprocessing errors should be equivalent. 

\begin{figure*}[]
\centering
\includegraphics[width=.9\textwidth]{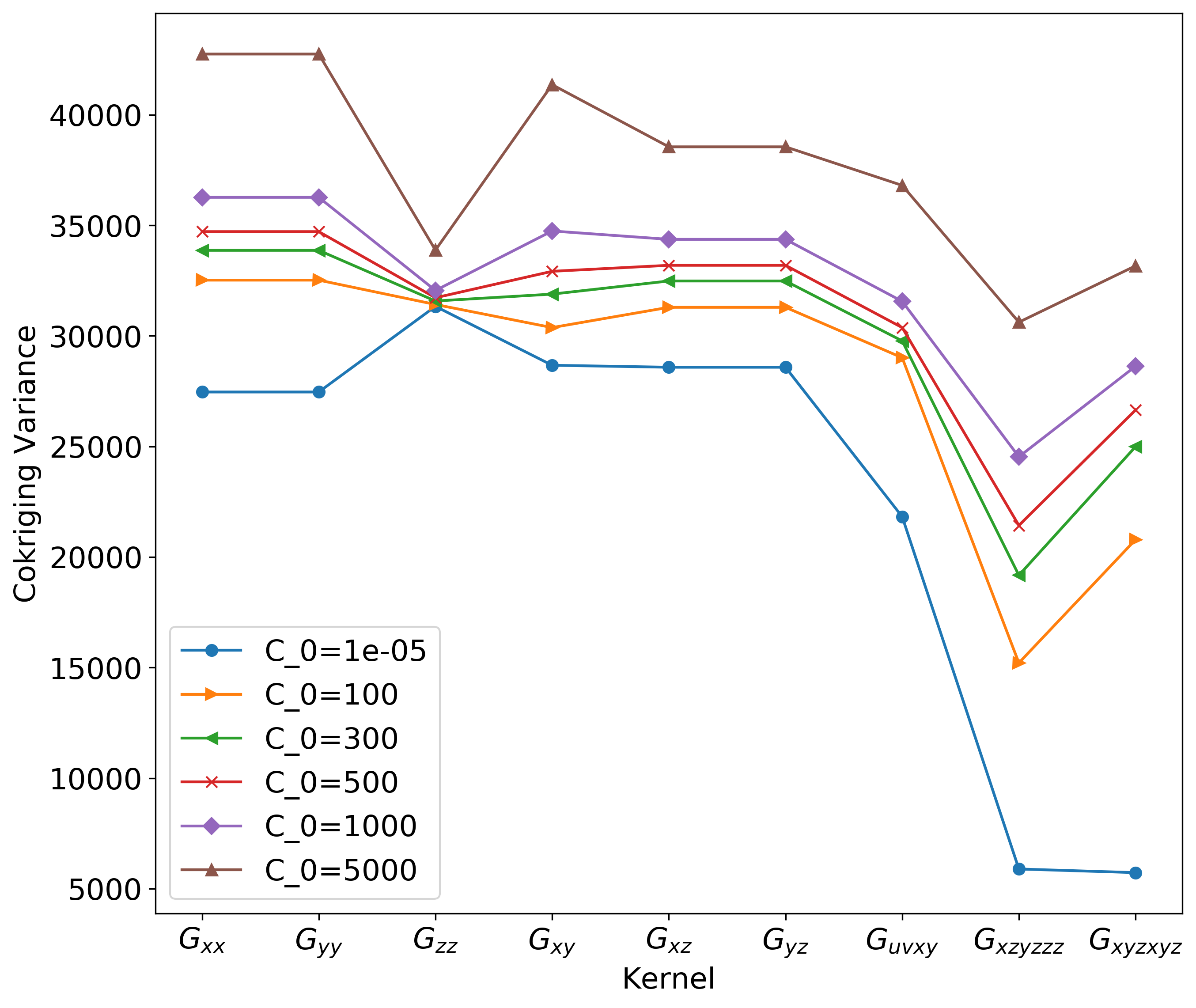}
\caption{Cokriging variances for several components and component combinations, as a function of increasing noise. Noise is added as a nugget effect.}
\label{fig:noise}
\end{figure*}

Using these weighted relative errors, the results are shown in Figure \ref{fig:noise}. As we expect, for all gradient components and combinations, the error variance increases by increasing noise. The component $Gzz$ and some of the component combinations show more resistance against noise increment. This figure also suggests that nugget noise increment, in general, increases the variance in the recovered models. 

\section{Case Study: New Found Dataset}
\begin{figure*}[]
\centering
\includegraphics[width=\textwidth]{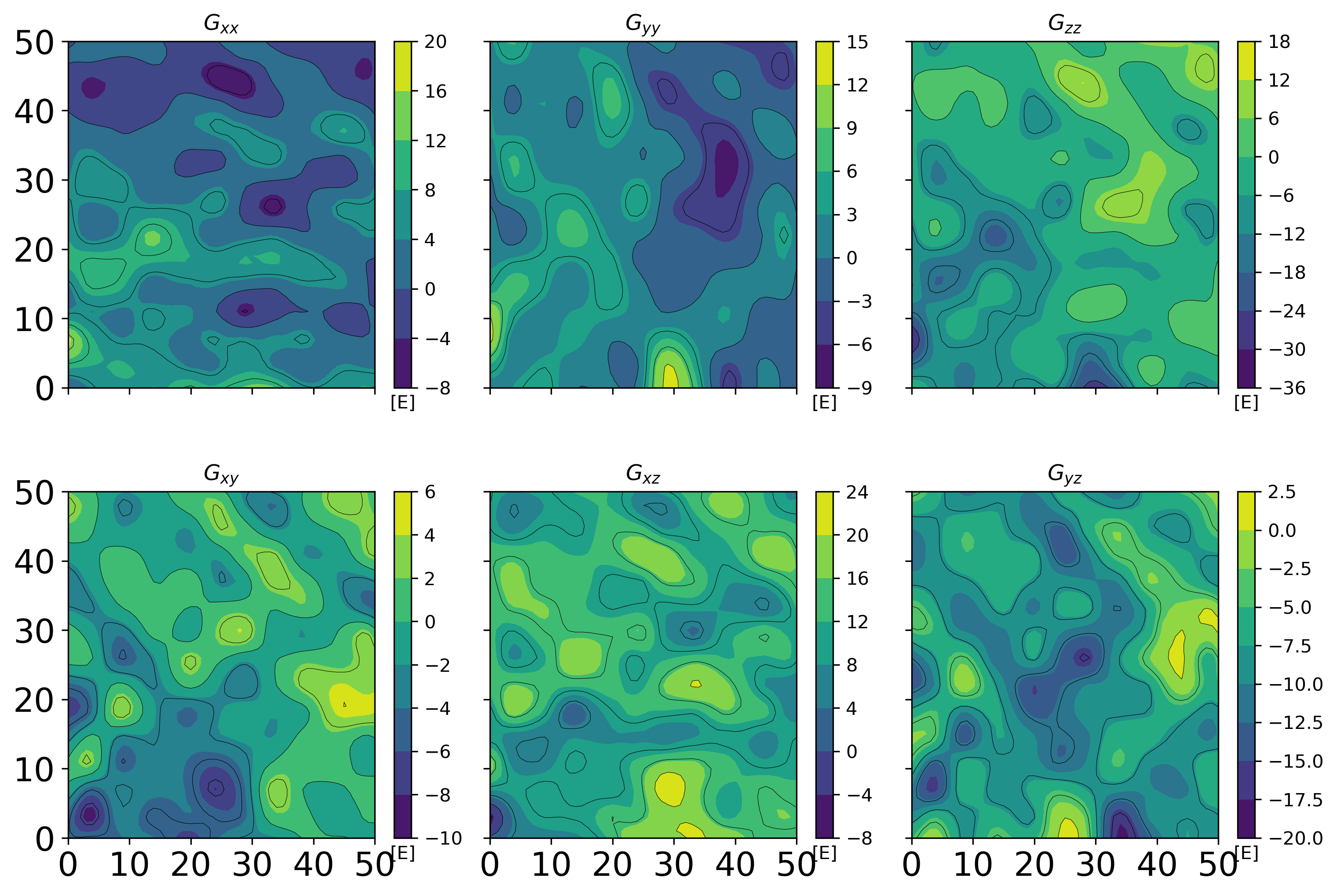}
\caption{Component quantities observed in a selected area in New Found dataset.}
\label{fig:newfound_observes}
\end{figure*}

Here we showcase the same inversion concepts and apply them to real-world solutions. In 2020, HeliFALCON Airborne Gravity Gradiometer (AGG) survey data was collected and provided by New Found Gold Corp. (NFGC)’s Queensway project in Newfoundland and Labrador, Canada, henceforth referred to in this paper as the New Found dataset. Of this data, we selected a $1 \times 1$ km area from the centre of this selection for use in this study. Using the stochastic inversion methodology as described above, we conduct the same analysis on this real-world data. Figure \ref{fig:newfound_observes}, shows the observed gravity gradient components for the selected area. 

\begin{figure*}[]
\centering
\includegraphics[width=.9\textwidth]{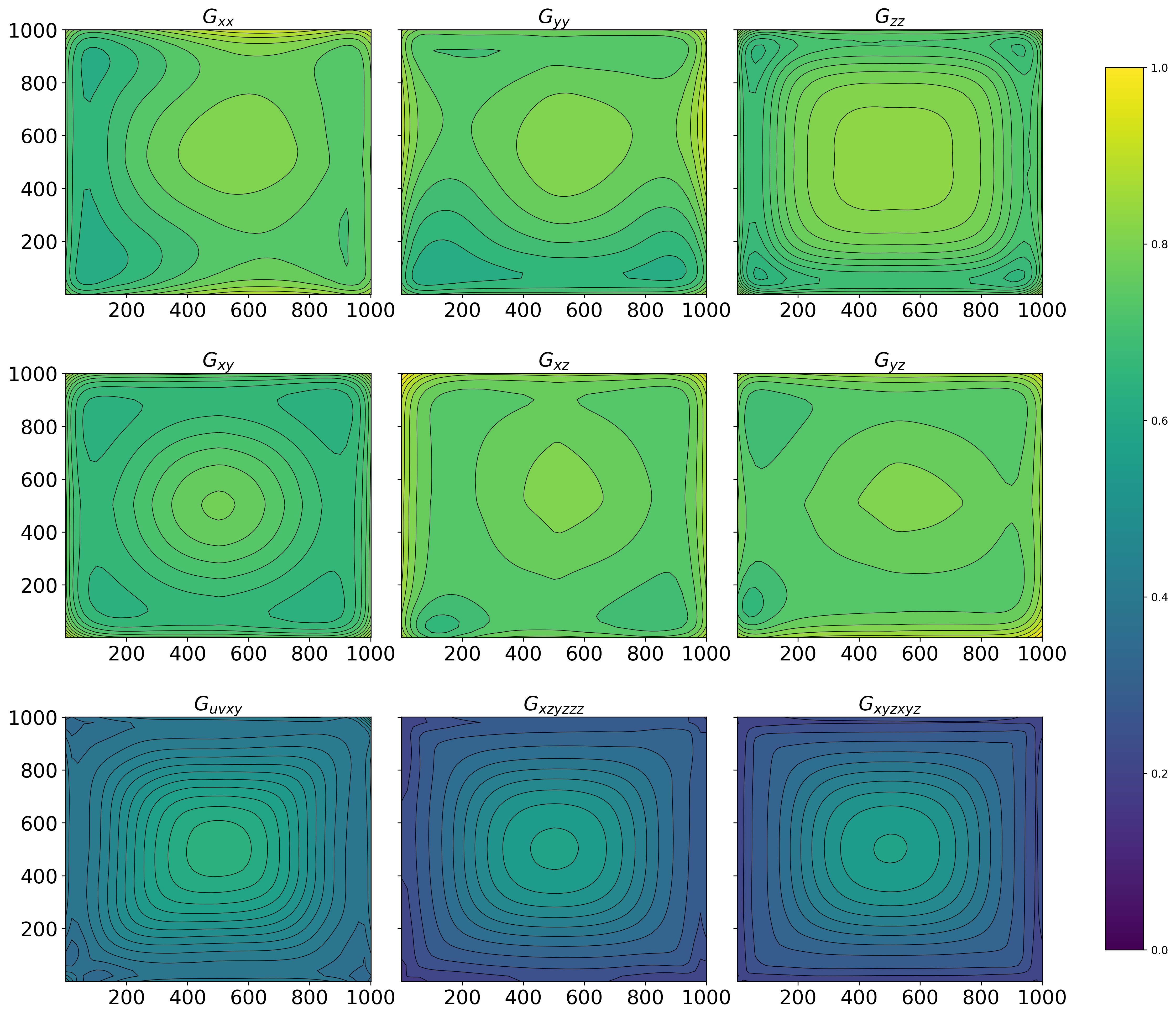}
\caption{Cokriging variances shown in section Z = 5 m depth for six components and three component combinations for the selected area in the New Found dataset.}
\label{fig:nf_variance}
\end{figure*}

Figure \ref{fig:nf_variance} shows the cokriging variances of different gravity kernels at the slice taken from a depth of 5 m. From these results we can clearly see the variance of a kernel combination is lower than that of a single component, most notably in the larger combinations. These results remain consistent with the observed results from the synthetic data in Figure \ref{fig:var_iso}.

\begin{figure*}[]
\centering
\includegraphics[width=.9\textwidth]{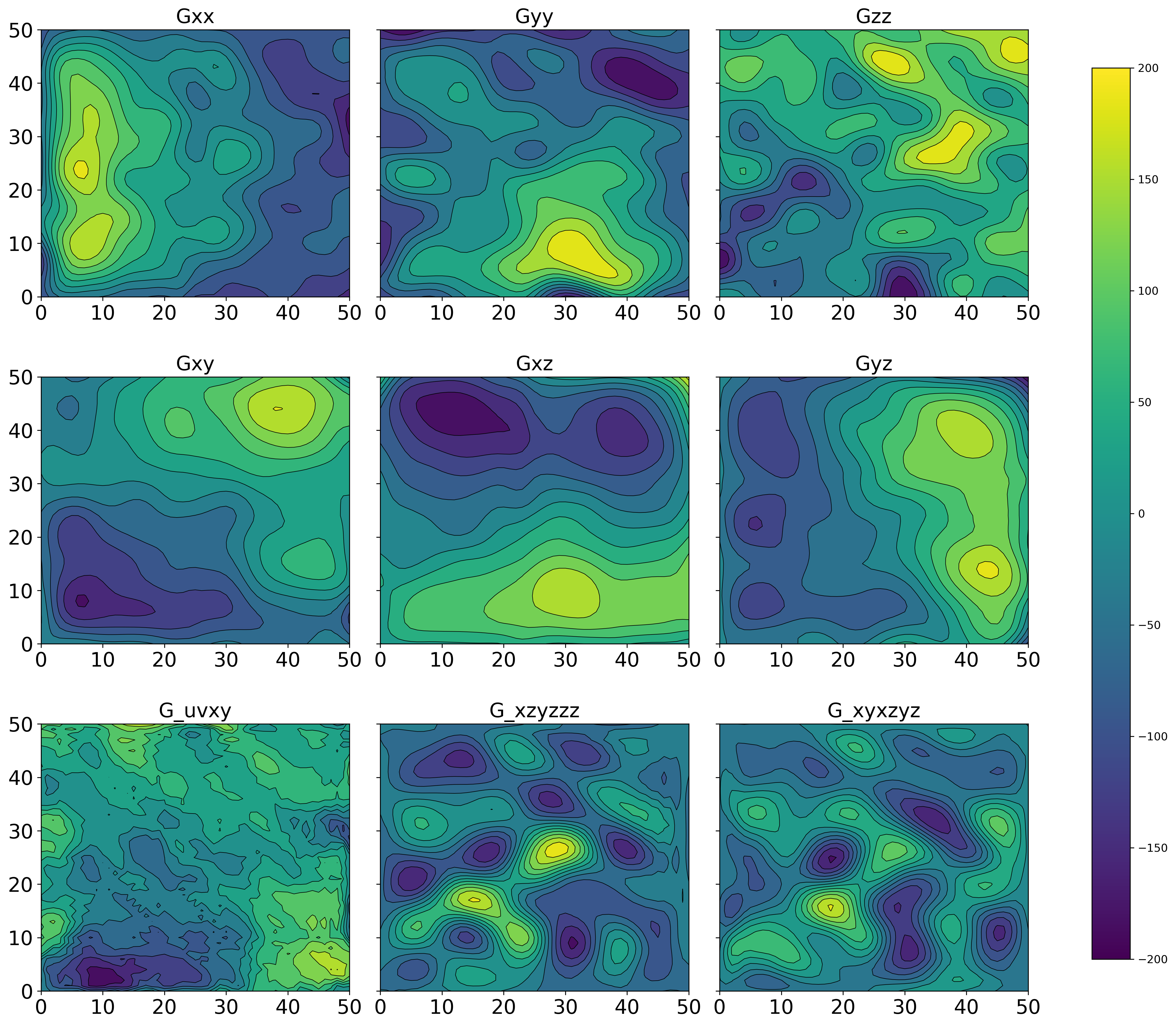}
\caption{The fifth layer of the recovered density model in Z direction on the selected area in the New Found dataset.}
\label{fig:nf_variance_ro}
\end{figure*}

Figure \ref{fig:nf_variance_ro} shows the density of prisms in the fifth layer below the surface in Z direction.

Using these stochastic inversion techniques, the generated density values can progress further through the processing pipeline for applications such as clustering and later structural interpretation.

%%---------------------------------------------Conclusion-------------------------------------------
\section{Conclusion} 
We used cokriging variances from stochastic inversion \citep{Shamsipour2010} as an indicator to quantitatively study different single tensor components. The indicator defines an error for each prism and the interpreter can easily select a single or combination of tensor components depending their purpose and available data. The method can simply be applied to study the effect of model geometry (in a stationary framework) or variation of the noise characteristics. For geometry studies, we analyze the effects of changing the variogram parameters: either ranges or rotations, as well as a study of the noise. 
We find that the combination tensor component errors are lower as compared to other components. We also find that error of combination tensor components remains independent of rotation, and continue to contain lower cokriging variances across all nugget effect values. These results remain consistent when the same inversion techniques are applied to the real New Found HeliFALCON AGG survey data.

\section{Acknowledgment}
We wish to thank Dr. Denis Marcotte for his valuable discussions throughout the preparation of this manuscript. The authors would also like to thank Dr. Mark Pilkington for his comments towards this work.

\bibliographystyle{elsarticle-harv.bst}
\bibliography{Gravity_Gradient}

\end{document}